\documentclass[fleqn,usenatbib]{mnras}

\usepackage[T1]{fontenc}

\DeclareRobustCommand{\VAN}[3]{#2}
\let\VANthebibliography\thebibliography
\def\thebibliography{\DeclareRobustCommand{\VAN}[3]{##3}\VANthebibliography}

\usepackage{graphicx}	
\usepackage{newtxtext,newtxmath}
\usepackage{orcidlink}

\title[Clumpy aerosols]{Resolving the flat-spectrum conundrum: clumpy aerosol distributions in sub-Neptune atmospheres}

\author[Owen \& Kirk]{James E. Owen$^{\orcidlink{0000-0002-4856-7837},1,2}$\thanks{E-mail: james.owen@imperial.ac.uk (JO)} \& James Kirk$^{\orcidlink{0000-0002-4207-6615},1}$
\\
$^{1}$Astrophysics Group, Imperial College London, Blackett Laboratory, Prince Consort Road, London SW7 2AZ, UK\\
$^{2}$ Department of Earth, Planetary, and Space Sciences, University of California, Los Angeles, CA 90095, USA
}

\date{Accepted XXX. Received YYY; in original form ZZZ}

\pubyear{2015}

\begin{document}
\label{firstpage}
\pagerange{\pageref{firstpage}--\pageref{lastpage}}
\maketitle

\begin{abstract}

Transmission spectroscopy of sub-Neptunes was expected to reveal their compositions and hence origins, yet many show flat near- to mid-infrared spectra. Such spectra can be explained either by metal dominated atmospheres or by high-altitude, grey aerosols. Observations of escaping hydrogen and helium from several of these planets rule out metal dominated atmospheres, while homogeneous distributions of small aerosols cannot produce flat spectra and large particles require unphysically high production rates. We investigate the role of heterogeneous, “clumpy” aerosol distributions in shaping transmission spectra. Modestly optically thick clumps at high altitudes can produce flat spectra even with small particles and physically realistic production rates. Clumping increases the effective photon mean-free path while reducing wavelength dependence, allowing the aerosol distribution to behave as an effective grey absorber. Applying this framework to the sub-Neptune TOI-776c, we show that clumpy aerosols can reconcile the observed flattening of its transmission spectrum with a primordial H/He-dominated atmosphere. We further discuss implications for emission spectra, where enhanced stellar radiation penetration and altered scattering in a clumpy medium could produce observable signatures. These results suggest that clumpy aerosol distributions naturally resolve the tension between flat spectra and low-metallicity atmospheres and may be a common feature of sub-Neptune exoplanets. More broadly, our results highlight the need to consider aerosol heterogeneity when interpreting high-altitude microphysics and the spectral appearance of exoplanet atmospheres with JWST, and motivate theoretical work to identify the physical mechanisms capable of generating clumpy aerosol distributions.

\end{abstract}

\begin{keywords}
planets and satellites: atmospheres --- planets and satellites: composition --- planets and satellites: individual: TOI-776c
\end{keywords}



\section{Introduction}

Exoplanet surveys have demonstrated that the most common type of exoplanets discovered to date are small ($\lesssim 4$~R$_\oplus$), low-mass ($\lesssim 20$~M$_\oplus$) planets on close-in orbits (periods $< 100$~days). This type of planet is found around most stars \citep[e.g.][]{Fressin2013,Dattilo2023}; however, the lack of Solar System analogues and the fact that they were not predicted by pre-{\it Kepler} planet formation models \citep[e.g.][]{Ida2008} mean their origins remain hotly debated \citep[e.g.][]{Bean2021}.

This class of close-in planets can be further subdivided. The radius valley represents a deficit of exoplanets with sizes of $\sim 1.8$~R$_\oplus$ \citep[e.g.][]{Fulton2017}, where the critical size is orbital-period \citep[e.g.][]{VanEylen2018} and stellar-mass dependent \citep[e.g.][]{Wu2019,VanEylen2021,Petigura2022}. Even before the radius valley was confirmed observationally, theoretical models predicted its existence \citep[e.g.][]{Owen2013,Lopez2013,Jin2014,Chen2016} as a transition between terrestrial composition planets and objects with about 1\% of their mass in primordial H/He atmospheres, arising from photoevaporation of their atmospheres. \citet{Weiss2014} and \citet{Rogers2015} used precise masses and radii for a population of planets to demonstrate that there was a transition in planetary composition at a radius similar to the radius valley. Planets with radii smaller than the (period-dependent) radius valley are consistent with an ``Earth-like'' bulk density, are likely terrestrial planets, and are often termed ``super-Earths''. Planets with larger radii must contain a voluminous volatile envelope/atmosphere to explain their bulk densities, and are often termed ``sub-Neptunes''. This result has only become stronger with increasing sample sizes and measurement precision \citep[e.g.][]{Parc2024}.  

The composition of sub-Neptunes’ volatile envelopes has been a matter of intense debate. Comparisons to mass-loss models suggest H/He-dominated atmospheres, irrespective of whether photoevaporation or core-powered mass-loss is considered \citep[e.g.][]{Owen2017,Gupta2019,Rogers2021}. These H/He-dominated atmospheres might then be enriched with heavy elements through chemical interactions and miscibility between the core and envelope \citep[e.g.][]{Schlichting2022,Misener2022,Rigby2024,Young2024,Bower2025,Heng2025,Werlen2025,Rogers2025c}, giving rise to higher-metallicity atmospheres. Alternatively, motivated by some planet population synthesis models, super-Earths and sub-Neptunes might form through different channels, with sub-Neptunes primarily forming outside the water snow-line before migrating to shorter periods \citep[e.g.][]{Zeng2019,Luque2022}. These formation scenarios, when combined with atmospheric escape, suggest that sub-Neptunes contain steam-dominated envelopes/atmospheres \citep[e.g.][]{Venturini2020,Burn2024}. Although detailed evolution models of multi-planet systems can break some of the degeneracies \citep[e.g.][]{OM16,OCE20}, at the population level the density distributions between ``water-worlds'' and H/He-rich sub-Neptunes are completely degenerate \citep[e.g.][]{Rogers2023}. While understanding the exoplanet population as a function of age might provide additional information \citep[e.g.][]{Rogers2025b}, direct measurement of the observable atmosphere's composition should provide a more direct probe of a sub-Neptune's envelope composition.  

Despite occasional successes \citep[e.g.][]{Madhusudhan2023,Piaulet2024,Benneke2024,Davenport2025}, recent JWST observations \citep[e.g.][]{Wallack2024,Cadieux2024,Alderson2025,Teske2025,Ahrer2025,Ohno2025} have demonstrated that the transmission spectra of many sub-Neptunes are flat (i.e. showing little or no evidence for wavelength dependence in their transmission spectrum) out to long wavelengths ($\sim$5~$\mu$m). These results imply that their atmospheres are either extremely metal-rich, or have optically thick, high-altitude aerosols (pressures $\lesssim$ 1 mbar) \citep[e.g.][]{Wallack2024}. However, there are challenges to both interpretations. Making the majority of sub-Neptunes have metal dominated atmospheres would make it difficult to match the location of the radius-gap with mass-loss models, pointing to a large (order-of-magnitude or larger) theoretical error in mass-loss model calculations \citep[e.g.][]{Owen2019}. This error would be difficult to reconcile with the success of the models in explaining direct observations of ongoing mass loss \citep[e.g.][]{dosSantos2023}. Furthermore, several of the planets -- TOI-776 b/c \citep{Loyd2025}, TOI-836c \citep{Zhang2025}  and GJ 3090 b \citep{Ahrer2025} -- have had primordial gas detected escaping from their atmospheres, indicating a large H/He reservoir in their envelopes. In the case of TOI-776 c the Lyman-$\alpha$ transit was detected with sufficiently high signal-to-noise to rule out a steam-dominated atmosphere as the source. Thus, if the flat-spectrum sub-Neptunes do indeed have H/He-dominated atmospheres, then the high-altitude aerosol solution is equally unsatisfactory. Producing high-altitude aerosols that are optically thick to $5+\mu$m requires large particles (with sizes larger than the typical wavelength of observation). However, large particles settle out of low-pressure environments extremely quickly. For example, the settling time for a 10~$\mu$m particle at 1 mbar in a typical sub-Neptune is only a few hours. Such large particle sizes would require aerosol production rates in excess of $10^{-9}$~g~cm$^{-2}$, significantly larger than those that can be produced by self-consistent microphysical models \citep[e.g.][]{Kawashima2019,Lavvas2019}.   

Therefore, we are left with a conundrum: either there is a missing piece of physics in our modelling of sub-Neptunes’ transmission spectra, or there must be a large theoretical error in our understanding of atmospheric composition and mass-loss, or of exoplanet aerosols. In this work, we explore the former. Instead of considering homogeneous aerosol distributions, which have been primarily used in previous modelling efforts, we consider the impact of a highly heterogeneous aerosol distribution. Instead of uniformly mixing aerosols within the atmosphere, we consider the impact of a ``clumpy'' aerosol distribution, in which small particles are concentrated into high-density regions separated by regions with low aerosol densities, similar to how cumulus clouds are common in Earth's atmosphere \citep[e.g.][]{Fowler1976}.

 In section ~\ref{sec:sketch} we present a simple introduction to how a heterogenous, or ``clumpy'' aerosol distribution results in transmission that is less sensitive to wavelength, using simple toy models. In section~\ref{sec:clumpy}, we introduce the basics of an effective approach to radiative transfer through a clumpy medium. In section~\ref{sec:detailed} we apply our model to TOI-776c a flat spectrum sub-Neptune with a large detected hydrogen exosphere. In section~\ref{sec:discuss}, we discuss our model in a wider context, including emission, along with comparisons to ``patchy'' clouds and fractal aerosols, before summarising in section~\ref{sec:summary}.

\section{Sketch of the idea} \label{sec:sketch}

We shall assume for the rest of this work that sub-Neptunes contain primordial hydrogen-dominated atmospheres (which could be enriched with heavy elements through formation or evolutionary processes). Therefore, explaining their generally flat, low-amplitude JWST NIR–MIR spectra requires high-altitude aerosol particles. As we shall explicitly demonstrate in Section~\ref{sec:detailed}, using large particles (which behave like grey absorbers when the wavelength, $\lambda$, is smaller than the particle’s radius, $a$) results in the need for implausibly high aerosol densities. Therefore, we are restricted to using small particles that satisfy $a\lesssim\lambda$. Small particles have a strong wavelength dependence, where the cross-section generically behaves as (for $a<\lambda$):
\begin{equation}
    \sigma \propto \left(a/\lambda\right)^{1+b} \label{eqn:cross_section_simple} 
\end{equation}
where $b$ is the Rayleigh index and is typically $>0$. For an absorber whose concentration scales with pressure as $X\propto P^{l}$, the transmission spectrum with respect to some reference wavelength ($\lambda_{\rm ref}$) is approximately given as \citep[][combining their equations S17 \& 7, and computing the relative transit radii as in \citealt{Heng2017}]{deWit2013}:
\begin{equation}
    \Delta R_T = \frac{H}{1+l}\log\left(\frac{\sigma_{\lambda}}{\sigma_{\lambda_{\rm ref}}}\right) \label{eqn:transit_depth1}
\end{equation}
Therefore, for aerosols with a uniform concentration ($l=1$), the transmission spectrum obeys the following:
\begin{equation}
    \frac{\Delta R_T}{H} = \left(1+b\right)\log\left(\frac{\lambda}{\lambda_{\rm ref}}\right)
\end{equation}
 Considering the wavelength range typically used for most JWST observations to date — the G395H bandpass (2.9–5.1~$\mu$m) — and assuming a typical sub-Neptune with a radius of 2.5~$R_\oplus$, a mass of 5~$M_\oplus$, and an equilibrium temperature of 800~K, we estimate the change in transit depth for $b=1$ around M- and K-dwarf hosts to be approximately 50–100~ppm over the bandpass. Such a signal would be detectable for typical targets. 
Thus, unless the aerosol optical depth is extremely high everywhere, it is challenging to obtain transmission spectra that lack a strong wavelength dependence when small aerosol particles are present. This condition would require the atmosphere to be optically thick at all wavelengths within the production region, which must be significantly smaller than the planetary scale height $H$ (i.e., $l \gg 1$).


However, the above analyses assume that the aerosols are homogeneously distributed through the atmosphere on scales smaller than a scale height. This need not be the case: for example, cloud coverage on Earth is non-uniform \citep[e.g.][]{Chen2000}. Not only is Earth’s cloud coverage known to be non-uniform, but on microstructural scales ($\lesssim H$), it has long been established that individual line-of-sight optical depths are often stochastic, with high variability on small scales through clouds \citep[e.g.][]{Nakajima1995}. Furthermore, in the case of the banded cloud structure seen on Jupiter, analysis indicates that the bands show sub-structure and significant variability in the cloud optical depths within individual bands \citep[e.g.][]{Matcheva2005,Fletcher2016}. Radiative transfer through highly heterogeneous and clumpy media fundamentally changes the resulting transmission, absorption, and reflection of the medium. \citet{Avaste1974} studied radiative transfer through ``broken'' clouds by developing a Markov process approach in a two-phase medium (cloudy and cloud-free), demonstrating that the clumpy nature allowed increased propagation of diffuse light, at the cost of wavelength dependence, behaving in a more grey manner. Motivated by the idea that highly heterogeneous aerosol distributions can give rise to weaker wavelength dependence -- and are the norm on Earth -- we will study the impact of a “clumpy” aerosol distribution on exoplanets.  

To intuitively understand why a clumpy aerosol distribution leads to weaker wavelength dependence, consider a single spherical clump of radius $R_c$. The column density from the centre to the edge of the clump behaves as:
\begin{equation}
    \mathcal{N} = n_c R_c \propto \left(\frac{M_c}{m_a}\right) \times \left(\frac{1}{R_c^2}\right) \label{eqn:col}
\end{equation}
where $M_c$ and $m_a$ are the mass of the clump and an individual aerosol particle, respectively.  As the size of a clump with fixed mass decreases, its volume shrinks faster than its surface area, leading to an increase in the column density through the clump.  

Consequently, compacting a fixed mass of aerosols into smaller, denser clumps can greatly enhance the optical depth {\it through the clumps}. If the clumps become sufficiently small, they can be optically thick even at long wavelengths, producing transmission spectra that are largely wavelength-independent. Because the effective path length in transmission is, by definition, marginally optically thick---with a characteristic length scale of $\sim \sqrt{2\pi R_p H}$ (\citealt{Fortney2005}), where $R_p$ is the planetary radius---even moderately sized clumps, with scales somewhat smaller than $H$, would be optically thick under these conditions.



\begin{figure*}
    \centering
    \includegraphics[width=\textwidth]{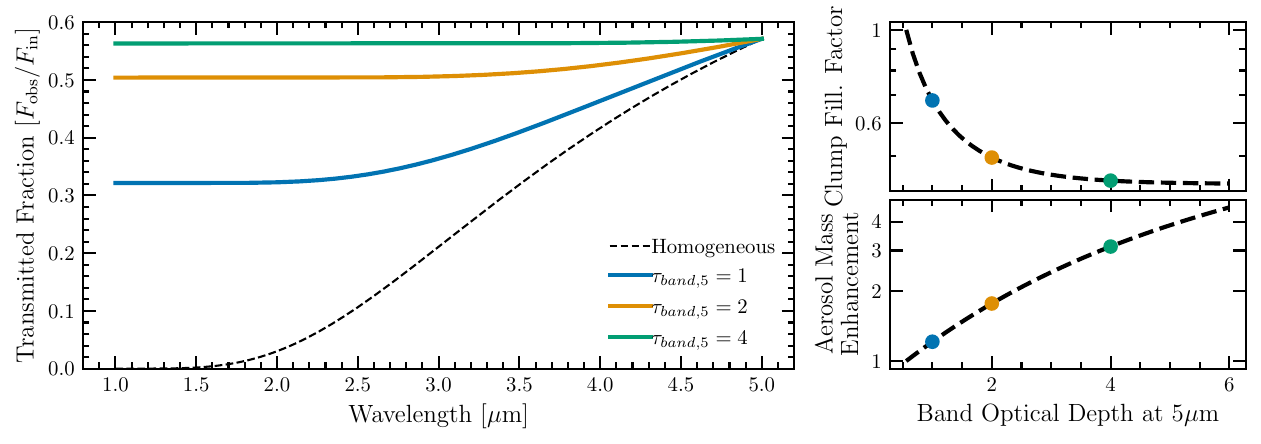}
    \caption{The left panel shows the fraction of transmitted flux through the simple box as a function of wavelength for our homogeneous model and simplistic “banded” model. As the optical depth through a band at 5~$\mu$m increases, the wavelength dependence of the transmitted flux decreases. The clumpy nature of the medium results in weaker wavelength dependence but higher transmitted flux due to the aerosol-free regions. The right-hand plots show how the filling factor ($N_{\rm bands}\ell/L$) and the total aerosol mass in the box (scaled to the homogeneous case) evolve with the band optical depth. As the optical depth through a single band increases, the total mass in aerosols in the box must increase to compensate. With a moderate increase in aerosol mass, a clumpy aerosol distribution can give rise to grey transmission, even with small particles.}
    \label{fig:simple_1}
\end{figure*}

\subsection{A toy demonstration}
To explicitly demonstrate the idea, consider a square box of volume $L^3$. We fill the box with a density of small (such that Equation~\ref{eqn:cross_section_simple} applies at all wavelengths of interest), but uniformly distributed aerosol particles, with number density $n_{\rm avg}$, where the density is defined such that the optical depth at 5~$\mu$m is $\tau_{5~\mu{\rm m}} \sim 0.56$\footnote{ The standard value for the transmission photosphere in an isothermal atmosphere with constant gravity is $e^{-\gamma}\approx 0.56$, where $\gamma$ is the Euler–Mascheroni constant, \citep[e.g.][]{Heng2017}.}. For this simple scenario, the optical depth through the box is trivially:
\begin{equation}
    \tau_\lambda = \tau_{5~\mu{\rm m}} \left(\frac{\lambda}{5~\mu{\rm m}}\right)^{-(1+b)}
\end{equation}
The spectrum measured by the observer (ignoring forward scattering) is, in terms of the ratio of the observed flux $F_{\rm obs}$ through the box compared to the incident flux $F_{\rm in}$ on the box, is given by:
\begin{equation}
    \frac{F_{\rm obs}}{F_{\rm in}} = \exp\left(-\tau_{5~\mu{\rm m}} \left(\frac{\lambda}{5~\mu{\rm m}}\right)^{-(1+b)}\right)\approx 1.-0.56\left(\frac{\lambda}{5~\mu{\rm m}}\right)^{-(1+b)}
\end{equation}
where we have used the small optical depth limit in the last equation. This demonstrates a strong wavelength dependence as expected, and as unwanted, given the observations.  

Now, consider the case where the density along the line of sight is increased for some rays while set to zero along others. In the most simple case, we can consider distributing the aerosols in $N_{\rm bands}$ of thickness $\ell$ with aerosol density $n_{\rm band}$ within the bands, leaving a fraction of $1-N_{\rm bands}\ell/L$ of lines of sight completely unobscured. The density within the bands is then adjusted to ensure that 56\% of the incident light is absorbed within the box at 5~$\mu$m, identical to the homogeneous case. The optical depth through a band is:
\begin{equation}
     \tau_{\rm band,\lambda} = \tau_{5~\mu{\rm m}} \left(\frac{n_{\rm band}}{n_{\rm avg}}\right)\left(\frac{\lambda}{5~\mu{\rm m}}\right)^{-(1+b)}
 \end{equation}
and the spectrum measured by the observer is:
 \begin{eqnarray} \label{eqn:clumpy_simple}
 \frac{F_{\rm obs}}{F_{\rm in}} &=& \left(1 - N_{\rm bands}\frac{\ell}{L}\right)\\ \nonumber &+& N_{\rm bands}\frac{\ell}{L}\exp\left(- \tau_{5~\mu{\rm m}} \left(\frac{n_{\rm band}}{n_{\rm avg}}\right)\left(\frac{\lambda}{5~\mu{\rm m}}\right)^{-(1+b)}\right)    
 \end{eqnarray}
In this toy problem, there are two controlling parameters. As we will show in Section~\ref{sec:clumpy}, these are the same parameters that appear in the full problem. The first is $N_{\rm bands}\ell/L \equiv V_{\rm ff}$, the volume filling factor of the bands. The second is $n_{\rm band}/n_{\rm avg}$, which quantifies the clumpiness of the medium and determines the column density through each band.
Equation~\ref{eqn:clumpy_simple} shows that as the clumpiness of the medium increases (by increasing $n_{\rm band}/n_{\rm avg}$), the flux transmitted through the layers decreases, and to maintain the fact that this must match $\exp(-0.56)$ at 5~$\mu$m, the volume filling factor must decrease. This increases the fraction of light transmitted through the aerosol-free regions, reducing any wavelength dependence. Taking into account the fact that the ratio of observed to incident flux must match $\exp(-0.56)$ at 5~$\mu$m yields, the analytic solution as a function of the volume filling factor:
\begin{eqnarray} \label{eqn:clumpy_simple2}
    \frac{F_{\rm obs}}{F_{\rm in}} &=& \left(1 - V_{\rm ff}\right)\\ \nonumber &+& V_{\rm ff}\exp\left(-\log\left(\frac{V_{\rm ff}}{{e^{-\tau_{5~\mu{\rm m}}}-(1-V_{\rm ff})}}\right)\left(\frac{\lambda}{5~\mu{\rm m}}\right)^{-(1+b)}\right) 
\end{eqnarray}

In the limiting case, the optical depth through the bands is effectively infinite and the volume filling factor of the bands is $\sim 43\%$, giving a perfectly wavelength-independent transmission profile.  

To demonstrate how a clumpy medium affects the wavelength dependence of transmission, we plot Equation~\ref{eqn:clumpy_simple2} in the limit between a homogeneous case $V_{\rm ff}=1$ and infinitely thick bands $V_{\rm ff}=1-\exp(-0.56)\approx 0.43$. The results are shown in Figure~\ref{fig:simple_1} for $b=1$, appropriate for tholin-like hazes, where we additionally show the fractional increase in total aerosol mass that occurs to maintain average transmittance, as well as the clump filling factor.  

\begin{figure}
    \centering
    \includegraphics[width=\columnwidth, trim=0.5cm 5.9cm 21.4cm 0, clip]{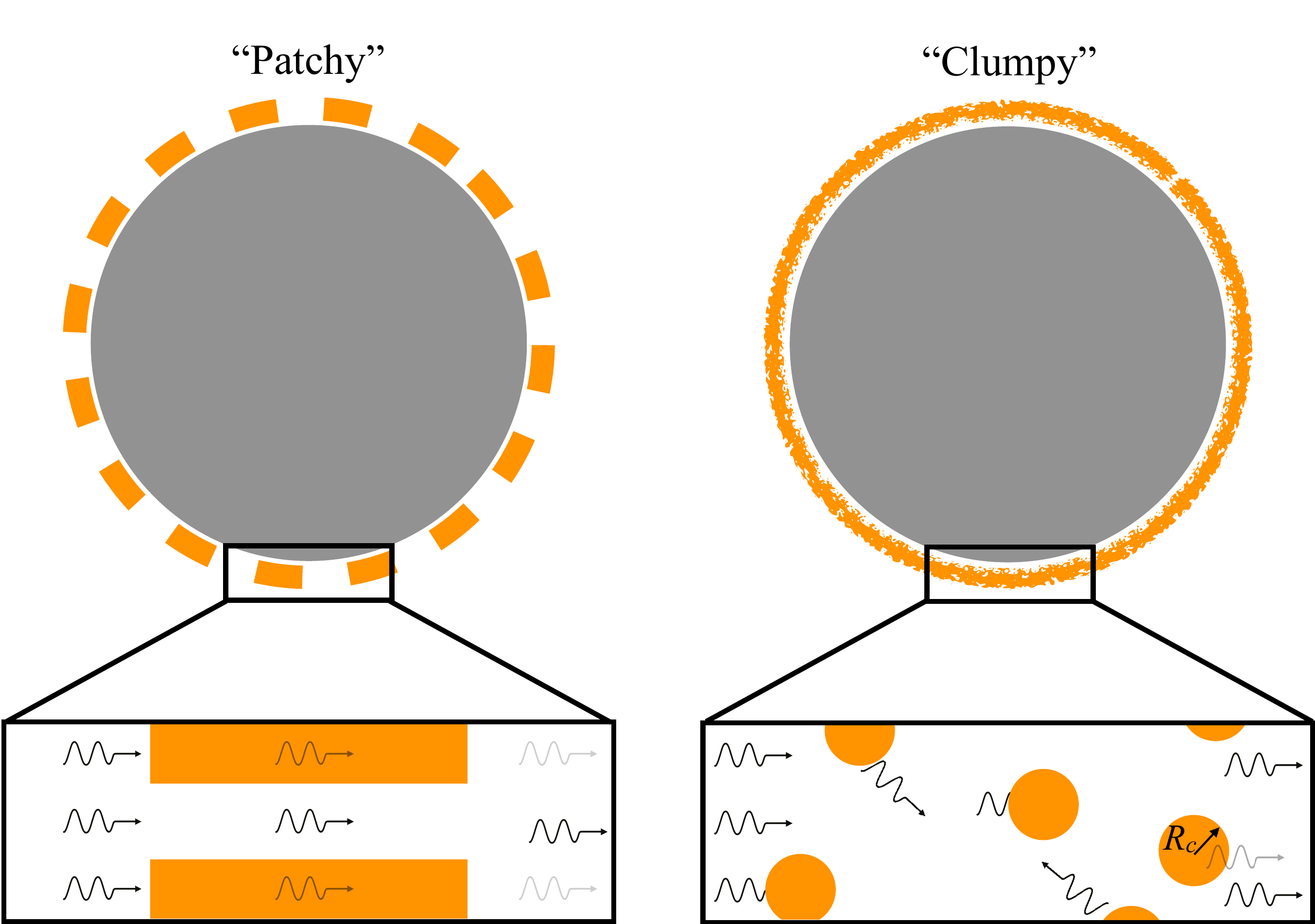}
    \caption{Schematic cartoon showing the difference between the patchy and clumpy formalisms, with a zoom in to the transmission geometry (where the star is to the left, and the observer to the right). In the patchy case, photons are either attenuated in the aerosol layers, or transmit freely (other than gas absorption) through the aerosol free regions. In the clumpy formalism, photons that do not interact with a clump pass freely through, uneffected (other than gas absorption). Photons that interact with a clump are generally scattered off their surfaces or completely absorbed, relatively few photons make it through a clump (in the case of optically thick clumps). The volume filling factor ($V_{\rm ff}$) of the clumps is the fractional volume of the region that contains aerosols particles.}
    \label{fig:cartoon}
\end{figure}

Since, by construction, an aerosol-dominated transmission spectrum is already marginally optically thick, only a small increase in local aerosol density (a factor of 2–3) is required to produce wavelength-independent transmittance with a clumpy density structure. Therefore, this simple experiment demonstrates the promise that a clumpy aerosol distribution has to flatten sub-Neptune transmission spectra. We note that this toy example of clumpy aerosol distributions suggests that we could treat them as ``patchy'' clouds, as implemented in some forward and retrieval models. However, as we shall demonstrate, such an analogy is not true in general, as scattering off clump surfaces and intra-clump absorption fundamentally changes the radiative transfer problem, modifying the effective single-scattering albedo of the medium, and cannot be modelled within the patchy formalism  (see Section~\ref{sec:patchy_discuss}). Therefore, while this example has been illustrative, one cannot generally use the patchy-cloud approach to model clumpy clouds as considered here, since they would predict the mean intensity, temperature, and hence observational properties incorrectly. This result becomes clear in the next section, where we demonstrate that the volume fill factor is not a fundamental independent variable in a general clumpy distribution, as the extinction properties are degenerate between the size of a clump and their volume filling factor.  We demonstrate the distinction between patchy and clumpy aerosol distributions in Figure~\ref{fig:cartoon}. 

\section{More realistic clumpy radiative transfer} \label{sec:clumpy}

The mathematical formalism to study radiative transfer through Earth's clumpy cloud distribution by \citet{Avaste1974} provided the basis for studies of transport through heterogeneous media in neutron transport \citep[e.g.][]{Malvagi1992} and star-forming regions in the ISM \citep[e.g.][]{Boisse1990}. As demonstrated with our simple analysis above, key results were that transmission through clumpy media became much less sensitive to wavelength, and the mean intensity inside the medium was significantly larger at high effective optical depth (due to multiple scattering inside the medium).  

The Markov process approach of \citet{Avaste1974} was studied in detail by \citet{Boisse1990} for transfer through a clumpy, dusty star-forming environment. \citet{Boisse1990} derived an effective homogeneous treatment for the optical properties of “clumpy” media when the volume filling factor of the clumps was small, and the scale of the clumps was small compared to the scale of the light travel path (in our case we require the clump size to be $\lesssim H$, the atmospheric scale height). \citet{Hobson1993} used \citet{Boisse1990}'s results to derive a simple approach by treating the clumps as large particles, showing a good match to the full Markov process approach in the small clump limit. Since this approach is intuitive, we shall adopt {their ``mega-grains'' approach here for our preliminary investigation in this work, noting that it is approximate. In the mega-grains approach, the {\emph{effective}} absorption ($\sigma_{\rm abs}$) and scattering ($\sigma_{\rm scat}$) cross-sections of the medium are given by:

\begin{equation}
    n_c\sigma_{\rm abs} = n_c\pi R_c^2\left(1-\omega_c\right)f_{\rm int}
\end{equation}

\begin{equation}
    n_c\sigma_{\rm scat} = n_c\pi R_c^2 \omega_c f_{\rm int}
\end{equation}

\begin{equation}
    \omega_c  =  \frac{\omega_{\rm aerosol}}{1+(1-\omega_{\rm aerosol})4\tau_c/3} \label{eqn:effective_SSA}
\end{equation}
where $n_c$ is the number density of clumps. $\omega_{\rm aerosol}$ is the single scattering albedo of the individual aerosol particles, and $\omega_c$ is the effective single scattering albedo of the clumps. $f_{\rm int}$ is the fraction of photons which enter a clump and interact with it, and $\tau_c$ is the optical depth from the centre to the surface of the clump (i.e. $\sigma_{a,\lambda}n_{a,c}R_c$, with $\sigma_{a,\lambda}$ the extinction cross-section of the individual aerosol particles, and $n_{a,c}$ the number density of aerosols inside a clump). All the expressions are functions of wavelength.  

\citet{Hobson1993} provide an expression for $f_{\rm int}$ for spherical clumps of:
\begin{equation}
    f_{\rm int} = 1 + \frac{\exp(-2\tau_c)}{\tau_c} - \frac{1-\exp(-2\tau_c)}{2\tau_c^2}
\end{equation}

Like the toy problem, mass conservation allows us to relate all the clump parameters in terms of the uniform case. As discussed above and by \citet{Boisse1990}, for a two-phase medium (high-density clumps and the low-density intra-clump medium) there are two free parameters that specify the radiative transfer problem: the mean extinction and the amplitude of the fluctuations. For our purposes, it is most intuitive to set the mean particle number density, $n_{\rm avg}$ (which does not depend on the properties of the clumps), since this is set by the aerosol production rate, and the column depth through a given clump $N_c$, from its centre to edge. Given these two, there is a degeneracy between the size of the clumps and their volume-filling factor ($V_{\rm ff}$). Thus, it is not necessary to choose either of these parameters. However, if one had knowledge of a physical mechanism that set, say, the clump radius, then the volume filling factor could be computed for that given optical depth. This degeneracy can be made clear where we can write the effective extinction as (assuming spherical clumps):

\begin{equation}
    \chi_{\rm eff} = n_c\left(\sigma_{\rm abs} + \sigma_{\rm scat}\right) = \frac{3}{4}\left(\frac{V_{\rm ff}}{R_c}\right)f_{\rm int}
\end{equation}
Now using the result that the column density through a single clump, from its centre to edge, is:
\begin{equation}
    N_c = n_{\rm avg} \left(\frac{R_c}{V_{\rm ff}}\right)
\end{equation}
we find that the effective extinction becomes:
\begin{equation}
    \chi_{\rm eff} = \frac{3n_{\rm avg}}{4N_c}f_{\rm int}
\end{equation}
We can interpret this result by noting that $\chi_{\rm eff}^{-1}$ is the effective mean-free path of photons through the clumpy medium. Therefore, we find that the effective mean-free path through the medium becomes:
\begin{equation}
    \chi_{\rm eff}^{-1} = \frac{4\tau_c \ell_{\rm hom}}{3f_{\rm int}}
\end{equation}
where $\ell_{\rm hom}$ is the photon mean-free path through a homogeneous medium with number density $n_{\rm avg}$.  Given that $f_{\rm int} \rightarrow 4\tau_c/3$ in the limit $\tau_c \ll 1$, we recover the homogeneous case in the optically thin regime, as expected. In the optically thick limit, $f_{\rm int} \rightarrow 1$ when $\tau_c \gg 1$, thus we obtain the following: 
\begin{equation}
    \chi_{\rm eff}^{-1} \rightarrow \frac{3 \tau_c\ell_{\rm hom}}{4}\quad{\rm when}\quad \tau_c\rightarrow\infty
\end{equation}
Essentially, at fixed average number density of aerosol particles, by clumping them we have increased the average photon mean-free path, as there are now optically thin intra-clump gaps through which the photons can propagate; however, by making an individual clumps optically thick we remove any wavelength dependence of the extinction: photons, irrespective of their wavelength, either freely propagate or are absorbed. We can demonstrate the impact on the effective Rayleigh index by computing the effective extinction due to a clumpy aerosol distribution:

\begin{equation}
    \sigma_{\rm eff} = \frac{\chi_{\rm eff}}{n_{\rm avg}}
\end{equation}

We determine the effective Rayleigh index between 0.5 and 5~$\mu$m as a function of the clump's column density (defined by its optical depth at 5~$\mu$m) for 0.1~$\mu$m particles of different compositions. We use the optical constants of soot \citep{Lavvas2017}, tholins \citep{Khare1984}, and silicates (Mg$_2$SiO$_4$, \citealt{Jager2003}) to calculate the radiative efficiencies for uniform spherical particles. To perform this calculation, we use the modified version of the {\sc bhmie} code \citep{Bohren1998} provided by the {\sc dsharp opac} package \citep[e.g.][]{Birnstiel2018}. We adopt a value for the internal density of the aerosol particles ($\rho_{\rm in}$) of $1~{\rm g~cm^{-3}}$ following \citet{Lavvas2011}, similar to the values found experimentally by \citet{Horst2013}. 

This calculation, shown in Figure~\ref{fig:simple}, shows that at moderate clump optical depths ($\gtrsim 0.3$), the slope of aerosol scattering can be sufficiently flattened. Therefore, similar to our toy model, with moderate clumping we suspect we can flatten sub-Neptune transmission spectra.  

\begin{figure}
    \centering
    \includegraphics[width=\columnwidth]{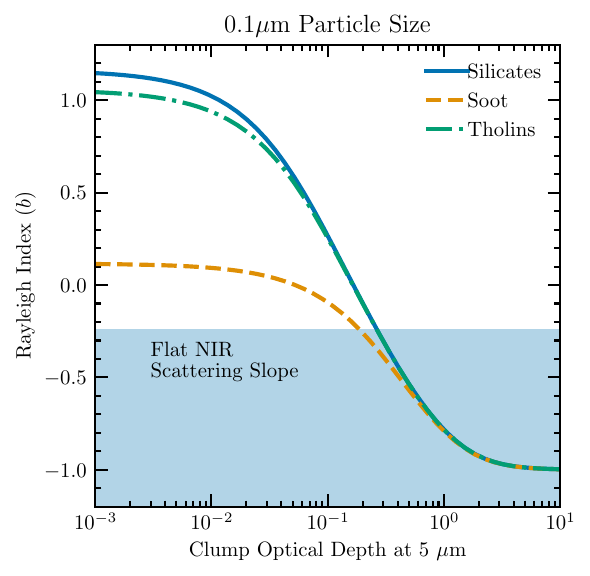}
    \caption{The Rayleigh index for 0.1~$\mu$m particle sizes as a function of clump optical depth at 5~$\mu$m for different aerosol compositions. The range of Rayleigh indices that would yield a flat aerosol spectrum for a TOI-776c-like planet at 20~ppm precision is shown as the blue shaded area. }
    \label{fig:simple}
\end{figure}

\section{A specific example: the case of TOI 776 c}\label{sec:detailed}

While there is now a growing sample of sub-Neptunes that all show flat spectra, TOI 776 c is an exemplar of this population. With a radius of $2.047$~R$_\oplus$, mass of $6.9$~M$_\oplus$, and a bulk density of $4.4^{+1.8}_{-1.6}~$g~cm$^{-3}$, it is inconsistent with an Earth-like composition at $>2\sigma$ \citep[e.g.][]{Fridlund2024}. Its bulk density requires a volatile-rich envelope, and it sits firmly in the degenerate region where it could have a steam-dominated atmosphere \citep[e.g.][]{Luque2021} or a H/He-dominated atmosphere \citep[e.g.][]{Rogers2023}. While this is true for many of the flat spectra sub-Neptunes, TOI-776c belongs to a growing category of these planets which also show evidence for an escaping primary atmosphere either in the form of helium \citep[e.g.][]{Ahrer2025, Zhang2025} or hydrogen \citep[e.g.][]{Loyd2025}. However, for TOI-776c the Lyman-$\alpha$ transits were sufficiently constraining that a water-dominated atmosphere could be definitively ruled out, since the observed neutral hydrogen cloud could not be produced by hydrogen from a dissociated steam atmosphere due to its measured size, speed, and temperature \citep[e.g.][]{Loyd2025}. Therefore, since TOI-776c requires a hydrogen dominated envelope to supply its Lyman-$\alpha$ exosphere, a likely explanation for its flat spectra is a high-altitude cloud deck ($P\lesssim 1$~mbar). However, as previously introduced, doing this is problematic, since the small particles generally expected to be present at low pressures do not act as the grey absorbers required to produce flat spectra in the G395H bandpass. In this section, we use specific models tailored to TOI 776c to demonstrate explicitly the problem with homogeneous cloud models, before demonstrating that a sufficiently optically thick clumpy aerosol distribution produces flat transmission spectra even with small particles. Indeed, as we shall show, it is easier to produce flat spectra with smaller aerosols in the clumpy cloud scenario, in contrast to the homogeneous cloud scenario, which requires large particles. While we purely model TOI-776c, we emphasise that the same approach will apply to the entire population of flat spectra sub-Neptunes.

\subsection{Methods}
The latest modelling of the interiors of sub-Neptunes suggests that some level of atmospheric enrichment in heavy elements is possible through envelope--core interactions. However, to make it as difficult for clumpy clouds to flatten the spectra as possible, we will use a Solar composition atmosphere, with a large scale height. Furthermore, the temperature range of sub-Neptunes that show flat spectra is where tholin-like hazes are predicted to be the dominant aerosols \citep[e.g.][]{Lavvas2017}. Therefore, we will use tholins as our aerosol particles  although we briefly explore soot-like hazes in Appendix~\ref{sec:soot}. 

\subsubsection{The JWST transmission spectrum of TOI-776c}

Two transits of TOI-776c were observed in May 2023 with JWST's NIRSpec/G395H instrument as part of the COMPASS programme (PIs: Batalha \& Teske, GO 2512). The analysis of these data and the planet's transmission spectrum was reported in \cite{Teske2025}. Here, we use the planet's spectrum as reduced with the \texttt{Tiberius} pipeline \citep{Kirk2017,Kirk2021}. 

Using chemical equilibrium forward models, \cite{Teske2025} found that the planet's featureless spectrum ruled out atmospheres $\lesssim 180-240\times$ solar metallicity, depending on reduction and modelling choices, for cloud-top pressures larger than 10$^{-3}$ bar. These metallicities correspond to mean molecular weights of $\sim 6-8$\,g\,mol$^{-1}$ and H$_2$O volume mixing ratios of $\sim9-13$\,\%. 
However, \cite{Teske2025} caution that mean molecular weight inferences are model dependent, highlighting some of the challenges associated with interpreting featureless spectra of small planets. 

\cite{Loyd2025} used their detection of Ly-$\alpha$ escape from TOI-776c to place an upper limit of 20\,\% on the planet's maximum atmospheric water content. Thus, while the atmospheric constraints from \cite{Teske2025} and \cite{Loyd2025} are consistent at the 3$\sigma$ level, in what follows we demonstrate that even lower metallicities ($1\times$ solar) are consistent with the JWST spectrum when considering our new clumpy aerosols distributions. 

\subsubsection{Computing Transmission Spectra}
To compute the transmission spectra with aerosols, we follow the approach in \citet{OMC25}. We use the {\sc petitradtrans} package to compute a chemical equilibrium atmospheric composition  (excluding condensation) between $10^2$ and $10^{-8}$~bar with 125 layers logarithmically spaced in pressure, assuming the atmosphere is isothermal at the equilibrium temperature. We use the H$_2$O line list from \citet{Water}, the CO line list from \citet{HITEMP}, the CH$_4$ line list from \citet{Methane}, the CO$_2$ line list from \citet{CO2} and collision-induced opacities from \citet{CIA}. We then use a modified version of {\sc petitradtrans} to compute the gas-only optical depths as a function of impact parameter. We do this at multiple reference pressures (defined as the pressure at a radius of 2.047~R$_\oplus$) from $10^{-8}$ to $10^{-1}$~bar. These models can then be combined with aerosols and interpolated to match the observed G395H white light radius.

Our individual tholin haze particles are parameterised in terms of their radius, $a$ (where we assume compact, spherical particles), and their production rate,  $\dot{\Sigma}_p$. Adopting the ballistic approximation, appropriate for the pressures of interest \citep[e.g.][]{Rossow1978,Ormel2019}, the terminal velocity approximation allows us to write the average haze density as a function of pressure as:
\begin{equation}
    n_{\rm avg} = \frac{\dot{\Sigma}_p}{t_{\rm stop} g m_a}
\end{equation}
where $t_{\rm stop}=a\rho_{\rm in}/v_{\rm th}\rho_{\rm gas}$ (with $v_{\rm th}$ the mean thermal speed of the gas and $\rho_{\rm gas}$ the gas' mass density) in the Epstein limit,  and $g$ is the planet's gravity. For the clumpy aerosol distribution, we choose the centre to edge optical depth of a clump at 5~microns to parameterise its column density ($\tau_{c,5}$). Therefore, given a particle radius, a haze production rate, and clump optical depth, we can calculate the effective optical properties of the clumpy clouds using the procedure described in Section~\ref{sec:clumpy}. 

We then compute the limb optical depths through the aerosol distribution assuming spherical symmetry at the terminator at the same impact parameters as our gas-only optical depths. We then combine the gas-only and aerosol optical depths and compute the wavelength-dependent transit radii using trapezoidal integration. This procedure is done within a root-finding scheme to match TOI-776c's G395H white light radius to a relative tolerance of $10^{-14}$ by adjusting the reference pressure. The stellar and planetary parameters used in our calculation are detailed in Table~\ref{tab:poseidon_params}. 

\begin{table}
    \centering
    \caption{The planetary and stellar parameters of TOI-776c used in our work, all taken from \citet{Fridlund2024}.}
    \label{tab:poseidon_params}
    \begin{tabular}{l|l}
         Parameter & Value\\ \hline

         \multicolumn{2}{l}{\textit{Planet parameters}} \\
         
         Planet radius ($\mathrm{R_P}$) & 2.047\,R$_{\oplus}$  \\
         Planet mass (M$_\mathrm{P}$) & 6.9\,M$_{\oplus}$  \\
         Equilibrium temperature (T$_{\mathrm{eq}}$) & 420\,K  \\ 

         \\ \multicolumn{2}{l}{\textit{Star parameters}} \\
         
         Stellar radius (R$_{\star}$) & 0.547\,R$_{\odot}$ \\
         Stellar mass (M$_\star$) & 0.542\,M$_{\odot}$  \\
         Stellar effective temperature (T$_{\mathrm{eff}}$) & 3725\,K \\
    \end{tabular}
\end{table}

\subsection{Homogeneous aerosol distributions} \label{sec:homogeneous}

\begin{figure*}
    \centering
    \includegraphics[width=\textwidth]{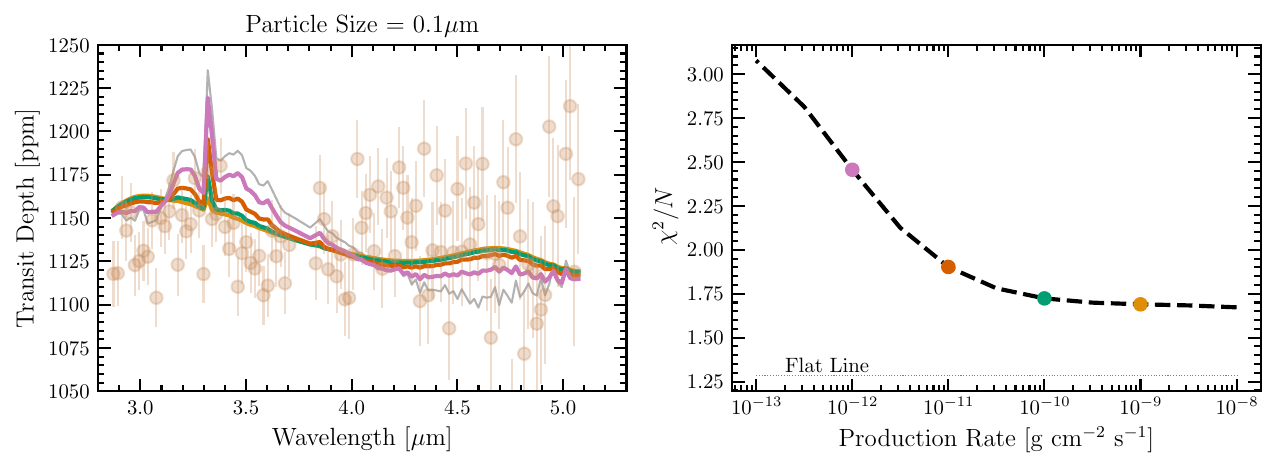}
    \includegraphics[width=\textwidth]{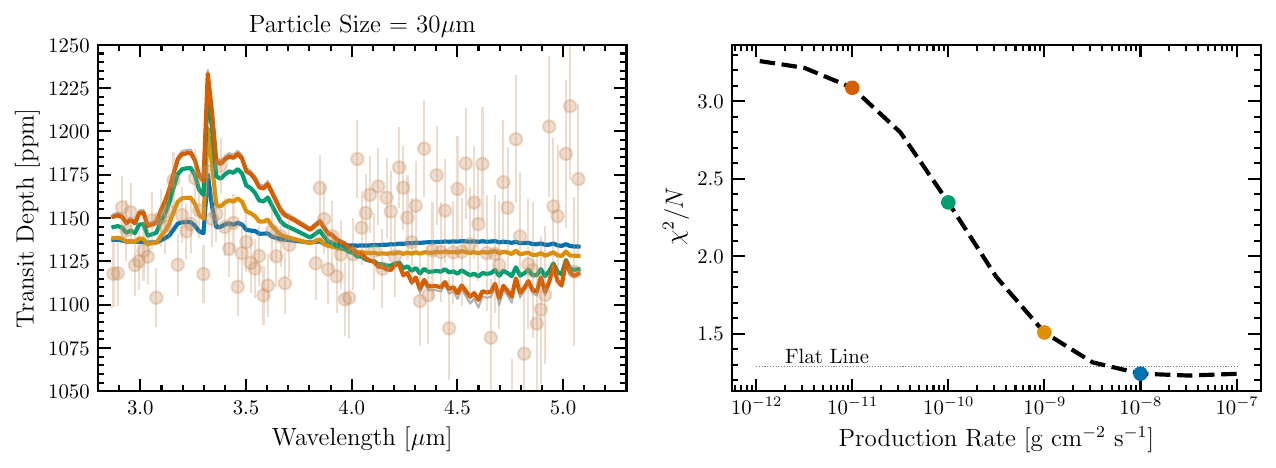}
    \caption{The left panels show the model transmission spectra of TOI-776c, where the thick coloured lines represent different production rates (indicated by the coloured points in the right panels). The thin grey line shows the aerosol-free transmission spectrum and the points show the observed G395H transmission spectrum from \citet{Teske2025}. The right panels show the $\chi^2/N$ goodness of fit metric as a function of the aerosol production rate, with the equivalent value for a flat line shown as the thin dotted horizontal line. The top row shows small aerosol particles, while the bottom row shows large aerosol particles. Small particles do not provide a good fit to the data at any production rate due to their strong wavelength dependence when $\lambda>a$. Large particles can provide a flat spectrum, but only at unphysically large aerosol production rates $\dot{\Sigma}_p \gtrsim10^{-8}~{\rm g~cm^{-2}}~{\rm s}^{-1}$. Therefore, it is challenging to explain sub-Neptunes' flat spectra with a homogeneous aerosol distribution in a primordial-dominated atmosphere.}
    \label{fig:homogeneous}
\end{figure*}

We first begin with an explicit demonstration of the challenge of fitting flat spectra with a homogeneous aerosol distribution. In this example, we replace the effective cross-sections with the true cross-sections of the aerosols. In Figure~\ref{fig:homogeneous}, we show the results for small, 0.1~$\mu$m, and large, 30~$\mu$m, haze particles for different production rates, spanning the nearly aerosol-free transmission spectrum to the completely haze-dominated case. We also show the $\chi^2/N$ goodness-of-fit metric used by the COMPASS program \citep[e.g.,][]{Wallack2024,Alderson2025,Teske2025} as a function of the production rate, comparing to the flat line value as a reference. 

These results clearly demonstrate the challenge of using a homogeneous aerosol distribution to explain flat transmission spectra at NIR to MIR wavelengths. Due to the wavelength dependence of small particles at $\lambda> a$, where they generally absorb less efficiently at longer wavelengths, they cannot produce a flat spectrum. For small particles, as the production rate increases, the spectrum tends to a slowly decreasing transit depth with wavelength, modulated by a broad tholin absorption feature at $\sim 4.7~\mu$m. This small-particle, aerosol-dominated spectrum is not a good fit compared to the data, even at extremely large production rates. For large particles that have constant absorption efficiencies with wavelength when $a\gg \lambda$, it is possible to produce flat spectra. Flat spectra require aerosols to dominate opacities at high enough altitudes so that molecular gas absorption is unimportant, i.e. pressures $\lesssim 1$~mbar \citep[e.g.][]{Wallack2024}. To counteract the fact that large particles settle extremely quickly (and for a fixed production rate larger particles have lower total surface area) one requires extremely high production rates, with flat spectra possible for production rates $\gtrsim 10^{-8}$~g~cm$^{-2}$~s$^{-1}$. Such production rates are one to two orders of magnitude higher than those found in self-consistent microphysical models, even in extreme cases \citep[e.g.][]{Kawashima2019,Lavvas2019}. 

Therefore, while the observed sub-Neptune flat spectra of those planets with escaping primordial atmospheres point to high-altitude aerosols, explaining them with homogeneous aerosol distributions appears challenging, as the expected small aerosols present at high altitudes produce a wavelength-dependent spectral slope at NIR to MIR wavelengths, while attempting to use large particles without such a wavelength dependence requires unphysically large production rates. 

\subsection{Clumpy aerosol distributions}

\begin{figure*}
    \centering
    \includegraphics[width=\textwidth]{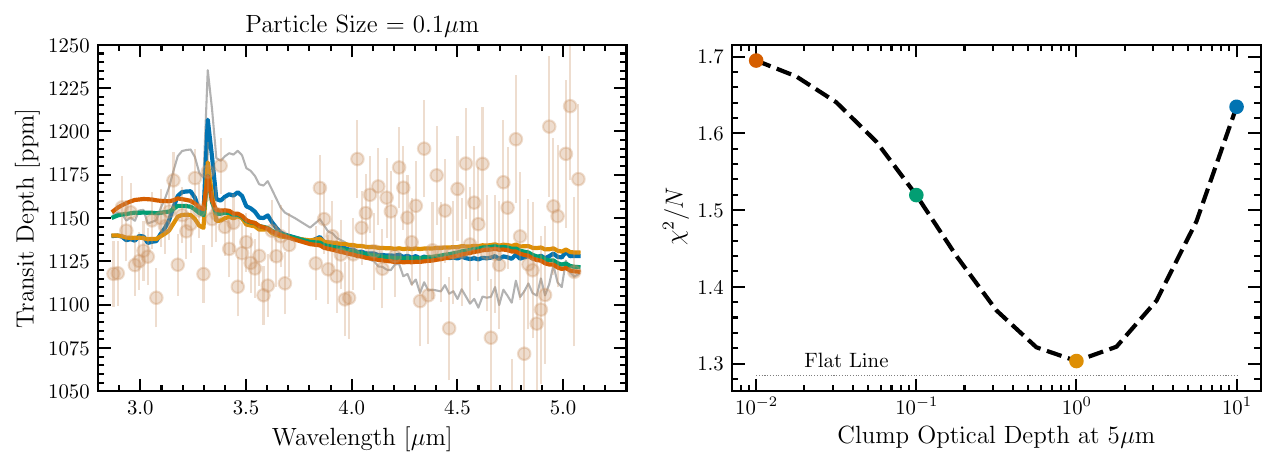}
    \includegraphics[width=\textwidth]{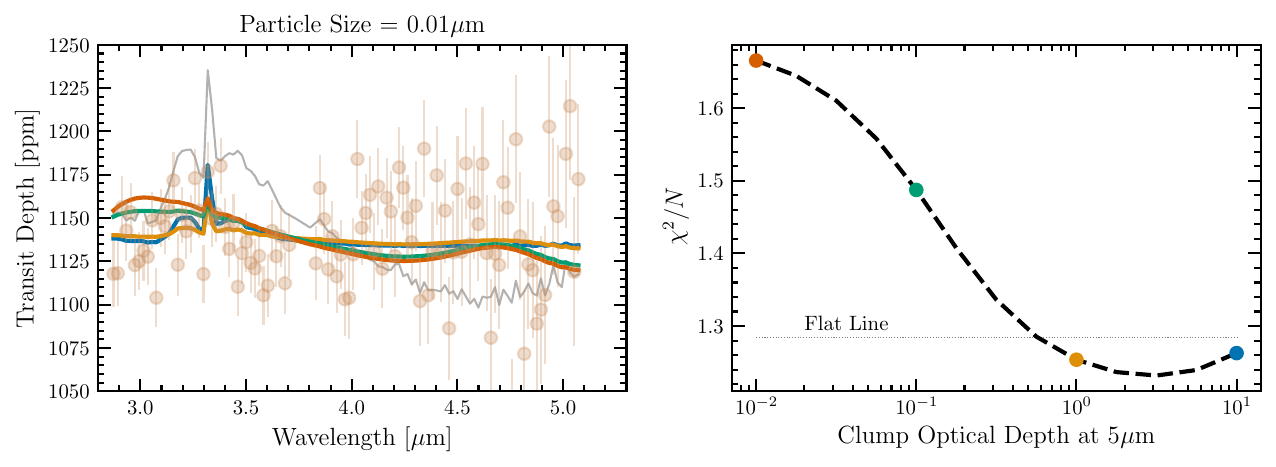}
    \caption{The left panels show the model transmission spectra of TOI-776c, where the thick coloured lines represent different clump optical depths at 5~$\mu$m (indicated by the coloured points in the right panels). The thin grey line shows the aerosol-free transmission spectrum and the points show the observed G395H transmission spectrum from \citet{Teske2025}. The right panels show the $\chi^2/N$ goodness-of-fit metric as a function of the clump optical depth, with the equivalent value for a flat line shown as the thin dotted horizontal line. The top row shows 0.01~$\mu$m particles, while the bottom row shows 0.1~$\mu$m particles. The production rate is $10^{-10}$~g~cm$^{-2}$~s$^{-1}$ in all models. These plots show that, provided the clumps are at least marginally optically thick at 5~$\mu$m, one can explain flat spectra with small particles and physical haze production rates.}
    \label{fig:clumpy1}
\end{figure*}

We now introduce a clumpy aerosol distribution, where we modify the optical properties of the aerosols to follow the effective approach for clumpy media. Given our earlier analysis, we expect that as the individual clumps become optically thick they will behave like grey absorbers; however, at fixed average number density, this will lead to a lower overall extinction opacity since, by clumping, the aerosols leave regions of the atmosphere which are aerosol-free. Thus, while we expect clumpy aerosol distributions to lead to flat spectra, they will not eliminate the need for aerosols to be present at low ($\lesssim$~mbar) pressures. 

In Figure~\ref{fig:clumpy1}, we show the model transmission spectrum for TOI-776c as a function of clump optical depth at 5~$\mu$m for a haze production rate of $10^{-10}$~g~cm$^{-2}$~s$^{-1}$. These show that, as the clump optical depth approaches unity, the amplitude of the $\sim3.3~\mu$m methane feature is significantly muted, and the entire spectrum becomes flattened, with a goodness of fit that approaches or, in some cases, exceeds that of a flat line. This begins to occur around a clump optical depth of 0.3, similar to that shown in Figure~\ref{fig:simple} for tholin-like hazes. As the clump optical depth continues to increase, the spectrum begins to approach that of the aerosol-free case. This is because, at high clump optical depths, all of the aerosols are effectively concentrated in a small number of high-optical-depth clumps. Thus, while the individual clumps are very optically thick themselves, acting almost like perfect absorbers, they do not contribute strongly to the overall opacity of the atmosphere, since they have a small covering fraction. Therefore, there is a preferred clump optical depth: too low and it tends towards the homogeneous case, too high and it tends towards the aerosol-free case, with a value around order unity being the preferred value to flatten spectra. We will return to this preferred value in Section~\ref{sec:discuss}.

 In Figure~\ref{fig:sizes}, we show the required clump sizes as a function of the clump volume filling factors for the models with a clump optical depth at 5$\mu$m of unity. These values have been computed in the transmission region. As highlighted in Section~\ref{sec:detailed}, there is a degeneracy between the clump sizes and their volume filling factors that produces the same clump column and hence effective extinction. However, this plot demonstrates that the clump sizes are smaller than the scale height for a range of plausible volume filling factors. This verifies the assumption that the aerosols are clumped on scales smaller than the scale height for this model, allowing the adoption of effective absorption and scattering cross-sections. 

\begin{figure}
    \centering
    \includegraphics[width=\columnwidth]{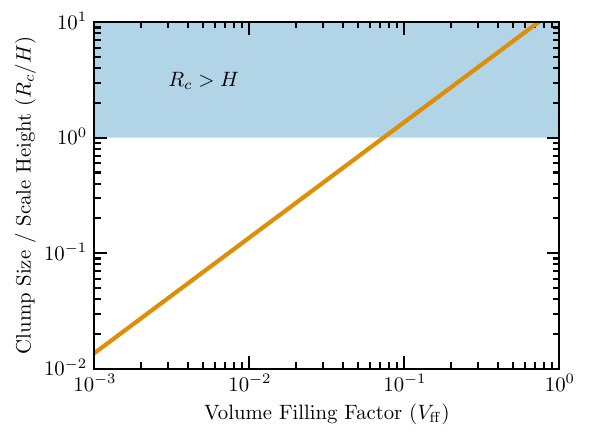}
    \caption{The implied clump sizes in the transmission region as a function of their volume filling factors for the $\tau_{c,5}=1$ model, with $a=0.1~\mu$m and $\dot{\Sigma}_p = 10^{-10}$ g cm$^{-2}$ s$^{-1}$ shown in Figure~\ref{fig:clumpy1}. We find clump sizes smaller than the scale height at moderate volume filling factors, validating our adoption of effective cross-sections.}
    \label{fig:sizes}
\end{figure}

Interestingly, Figure~\ref{fig:clumpy1} also demonstrates that the effect of particle size is opposite to that of the homogeneous case. Since smaller particles settle more slowly, higher haze densities are reached at high altitude for a fixed production rate. Therefore, it is easier to suppress the molecular gas features (methane in our example) and match the data with smaller particles, provided they can produce similar clump optical depths. 

We can analyse this in more detail by studying the spectrum as a function of the optical depth of the clump and the haze production rate for different particle sizes in Figure~\ref{fig:param}. This figure demonstrates that below a clump optical depth of $\lesssim 0.1$, the models approach the homogeneous case, as one would expect; while at higher clump optical depths, the spectra become flatter and a better fit to the data. As seen before, at extremely high clump optical depths, the spectrum tends towards the aerosol-free case, and the fit to the data again becomes poor. 

These calculations clearly demonstrate that flat spectra can be produced provided the clump optical depth is $\gtrsim 0.3$ at 5~$\mu$m and the aerosols distribution is optically thick at pressures $\lesssim 1$~mbar. High-altitude aerosols do require high production rates; however, unlike the homogeneous case, this can be mitigated with smaller particles. This allows the required production rates to fall well within the physical range predicted by the microphysical models ($\lesssim 10^{-9}$~g~cm$^{-2}$~s$^{-1}$). Therefore, at least phenomenologically, a high-altitude clumpy aerosol distribution appears to be a solution to the conundrum of how sub-Neptunes can have flat spectra measured by JWST in the 3–5~$\mu$m range and predominantly primordial atmospheres dominated by H/He as indicated by their exospheres. The corollary of our results is that current flat JWST G395H transmission spectra of sub-Neptunes do not provide any constraint on their atmospheric metallicity when accounting for the possibility that their aerosol distribution is clumpy. 

\begin{figure}
    \centering
    \includegraphics[width=\columnwidth]{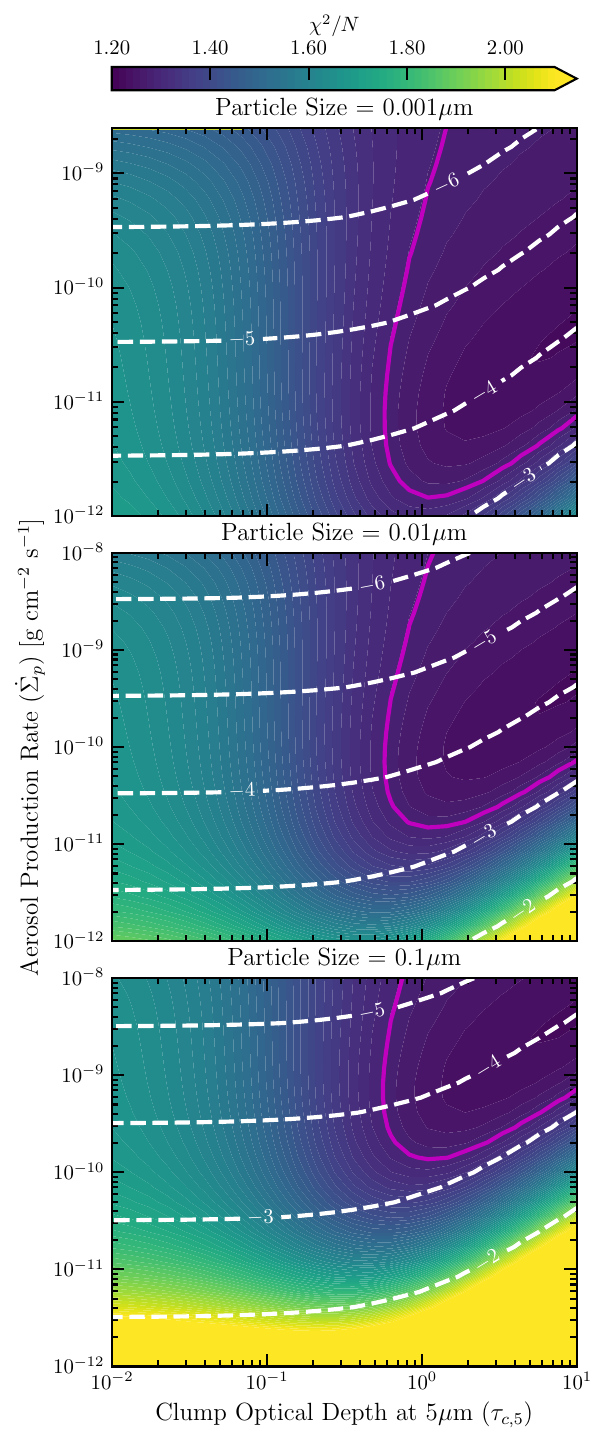}
    \caption{The goodness of fit to the TOI-776c G395H transmission spectrum of clumpy aerosol models as a function of the clump optical depth at 5~$\mu$m and the haze production rate for three different (small) particle sizes. The solid magenta contour shows the goodness-of-fit value for a flat line. The dashed white contours show the pressure values (in $\log_{10}$~bar) at which just the aerosol distribution has a photospheric optical depth in transmission (i.e. $\tau=e^{-\gamma}$).}
    \label{fig:param}
\end{figure}

\section{Discussion} \label{sec:discuss}

We have demonstrated, using phenomenological models, that a clumpy aerosol distribution can explain the observed flat JWST NIR to MIR transmission spectra of sub-Neptunes, without requiring unphysically large aerosol particles or high mean-molecular-weight atmospheres. Therefore, while helpful in explaining the observed flat spectra of sub-Neptunes whose exospheric measurements imply primordial H/He-dominated atmospheres, we have introduced an additional degeneracy into the modelling of sub-Neptune atmospheres.  

\subsection{Extrapolation to longer and shorter wavelengths}

\begin{figure*}
    \centering
    \includegraphics[width=\textwidth]{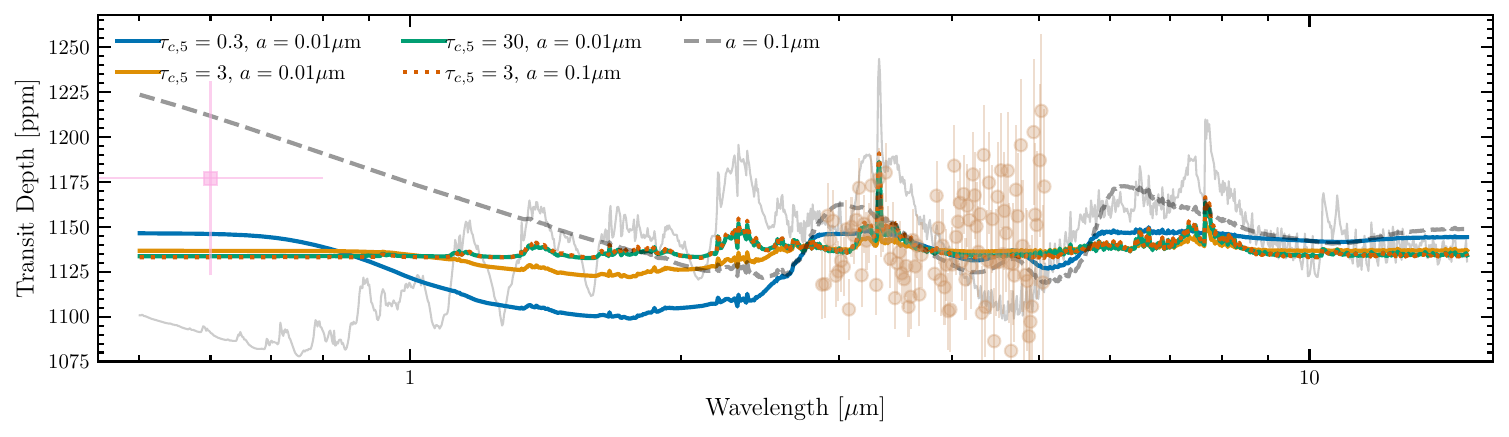}
    \caption{Modelled transmission spectra of TOI-776c from 0.5 to 15~$\mu$m. The coloured lines represent a clumpy aerosol model, while the dashed grey line is for a homogeneous aerosol distribution. The thin grey line shows the aerosol-free, solar abundance chemical equilibrium model. The data points are the observed G395H JWST spectrum from \citet{Teske2025} and the CHEOPS optical transit from \citet{Fridlund2024}. In all cases the production rate is $10^{-10}$~g~cm$^{-2}$~s$^{-1}$.}
    \label{fig:long_wav}
\end{figure*}

We now consider how the different models appear outside the G395H bandpass. Although this bandpass was originally considered an excellent option for studying exoplanet atmospheres because of its coverage of many major carbon- and oxygen-carrying molecules \citep[e.g.][]{Batalha2017,GuzmanMesa2020,Constantinou2022}, it is becoming increasingly clear that modelling degeneracies necessitates a wider wavelength coverage \citep[e.g.][]{Lueber2024}. Therefore, we model our transmission spectra at NIR wavelengths and longer in to the MIR by computing $R=400$ model spectra between 0.5 and 15~$\mu$m. As in the previous section, we ensure that these model spectra reproduce the white light radius of the JWST G395H spectrum. We plot both homogeneous and clumpy aerosol models in Figure~\ref{fig:long_wav}. In all cases, we set the production rate to a value high enough to be as good a fit as possible to the data in the G395H bandpass (noting, as detailed in Section~\ref{sec:homogeneous}, that the small, $a=0.1~\mu$m homogeneous model does not provide a good fit to the data), which is chosen to be $10^{-10}$~g~cm$^{-2}$~s$^{-1}$. 

With the exception of the homogeneous $a=0.1~\mu$m case, we find generally flat spectra over the full range of wavelengths at the $\lesssim 50$~ppm level. Interestingly, for marginally optically thin clumps (e.g. $\tau_{c,5}=0.3$), we do see a reduced transit depth in the 1–2.5~$\mu$m range. This is due to an opacity window in tholin-like hazes and is thus unique to this haze species (e.g. soot-like hazes or other cloud species will not show this drop, a result we demonstrate for soot-like hazes in Appendix~\ref{sec:soot})\footnote{Such a rise in transit depth with increasing wavelength is also characteristic of stellar contamination from unocculted active regions, which has been potentially observed \citep[e.g.][]{Kirk2021}.}. Therefore, we generally expect a clumpy aerosol distribution to produce a flat spectrum over the full optical–MIR wavelength range currently accessible. Thus, with the exception of marginally optically thin clumpy tholin-like hazes, a clumpy aerosol distribution would be degenerate with both a high mean-molecular-weight atmosphere and large aerosols, although, as discussed above, these latter scenarios appear to be unlikely. Therefore, additional wavelength coverage in transmission is unlikely to further break this degeneracy.   

\subsection{Impact on emission spectra} \label{sec:emission}

So far we have only discussed the impact of a clumpy aerosol distribution on transmission spectra. While clumpy aerosols naturally provide a solution to the conundrum of flat spectra at NIR to MIR wavelengths in primordial atmospheres, they have increased the size of the already degenerate parameter space. In fact, if marginally optically thick aerosol clumps are present in sub-Neptune atmospheres, transmission spectra are unlikely to provide insights into their compositions over the broad range of wavelengths currently accessible. 

However, by clumping the aerosols into optically thick regions, the effective extinction has dropped because of the creation of optically thin intraclump regions. This dramatically increases the ability of incident radiation to penetrate the atmosphere. \citet{Boisse1990} and \citet{Hobson1993a} both demonstrated that the mean radiative intensity could be significantly higher in a clumpy medium compared to a homogeneous one, with \citet{Hobson1993a} showing this increased as the number of phases with distinct densities increased. In turn, due to the clumpiness, the reprocessed thermal radiation can escape the atmosphere more easily (a process well established in terms of escape fractions from galaxies, e.g., \citealt{Neufeld1991}). Furthermore, as demonstrated in Equation~\ref{eqn:effective_SSA}, a clumpy medium has a lower single scattering albedo than a homogeneous one. This can be understood easily: instead of scattering off a single photospheric surface, scattering now takes place off clump surfaces, producing a more isotropic scattered radiation field, where scattered photons are more likely to encounter another clump and be absorbed.

Therefore, we can speculate that while transmission spectra might be degenerate, emission spectra are likely to be sensitive to any clumpy nature of the aerosol distribution. The larger mean intensity of stellar radiation deeper in to the day-side atmospheres of the planets will result in hotter temperatures, and the ease of escape will allow deeper emission photospheres to be reached. Both these effects are likely to lead to stronger emission features. Thus, a natural next step will be to include effective clumpy aerosol optical properties into self-consistent radiative transfer codes that are coupled to the atmosphere's thermal structure to consider the impact of clumpy aerosols on emission spectra. Such modelling would then facilitate planning observations of JWST targets that have already been observed with flat NIR to MIR transmission spectra.  

\subsection{What might create a clumpy aerosol distribution}

Throughout this work, we have approached modelling heterogeneous, clumpy aerosol distributions purely in a phenomenological way, parametrising their structure without reference to any physical mechanism that might generate such a distribution. Thus, the precise physical mechanisms that could produce clumpy aerosols in sub-Neptune atmospheres remain uncertain, although we can speculate about several possibilities.  

In established work, atmospheric dynamics, including strong vertical and horizontal winds, can create inhomogeneous aerosol structures \citep[e.g.][]{Smith1998}. Turbulent mixing is well established to lead to the aggregation of small aerosol particles into denser clumps, particularly at high altitudes where the mean free path of gas molecules is large \citep[e.g.][]{Grace2025}. Additionally, in the case of clouds, condensation and coagulation in regions of locally enhanced supersaturation may preferentially form clumps rather than a uniform aerosol layer \citep{carslaw2022aerosols}.  

We can also look to other astrophysical contexts. Two-fluid instabilities between solids and gas are well established in a wide range of environments \citep[e.g.][]{Squire2018}. The requirement to reach clump optical depths of order unity, without reaching much larger values, might point to radiative feedback. \citet{Lyra2013} identified such a clumping instability in debris disks between solids and particles that are weakly thermally and dynamically coupled, conditions likely present at low pressures in exoplanet atmospheres, and as such this is a prime candidate for a clumping mechanism. Furthermore, \citet{Owen2020_snow} identified a thermal condensation instability in protoplanetary discs that operates at optical depths of order unity. The radiative transfer geometry at a planet's terminator is similar to passively heated accretion discs, as such, this mechanism should be explored in the context of cloud condensates.  

Further dynamical investigations should be performed on small scales to identify and study any possible clumping mechanisms that might operate at high altitudes in sub-Neptune atmospheres.

\subsection{Comparisons with alternative approaches}
Previous work has shown that both fractal or ``fluffy'' aggregates and ``patchy'' aerosol distributions can lead to flattening of transmission spectra. It is worth discussing the similarities and differences between clumpy aerosol distributions and these suggestions. 
\subsubsection{Fractal aggregates}
As an alternative to clumpy aerosol distributions, fractal aggregates could potentially provide a similar effect. Fractal aggregates have good experimental support, which shows that particles can grow into aggregates with high porosity and low internal density in low pressure environments \citep[e.g.][]{Blum2000}. Fractal aggregates increase the effective extinction opacity of a particle \citep[e.g.][]{Lodge2024}, while slowing their settling \citep[e.g.][]{Sorensen2001}. \citet{Adams2019} demonstrated that fractal aggregates could flatten the transmission spectra of hot Jupiters without appealing to large particle sizes or high production rates. Similarly, \citet{Lavvas2019} \& \citet{Ohno2020b} demonstrated that fractal aggregates could produce flatter spectra for GJ1214 b compared to compact spheres, although they both did find a spectral slope. Additionally, they both argued that the  flatness found by \citet{Adams2019} was a results of their specific choice of fractal dimension and that other choices resulted in detectable spectral slopes.  In terms of transmission, fractal aggregates and clumpy aerosol distributions essentially do similar things: reduce the wavelength dependence of their extinction opacity \citep[e.g.][]{Moran2025}. In a specific application to GJ1214 b for KCl clouds, \citet{Moran2025} found that fractal aggregates for a range of parameters would yield a flat optical transmission spectrum; however, spectral features were still present at 3-5$\mu$m. 

Thus, fractal aggregates and clumpy aerosol distributions can each flatten transmission spectra. However, whether homogeneous fractal aerosols alone can do so for sub-Neptunes in the 3–5 $\mu$m G395H range is still uncertain, as they would need to achieve this within primordial H/He atmospheres to explain the exospheric detections. Even if fractal aggregates and clumpy aerosols do indeed turn out to be degenerate in transmission, in emission, one can speculate that they might behave differently. As discussed in Section~\ref{sec:emission}, clumpy aerosol distributions effectively behave like aerosol distributions with a lower single scattering albedo, while increasing the mean intensity of the penetrating stellar radiation field. Alternatively, coherent scattering inside a fractal aggregate can increase the single-scattering albedo of the particles over that of a single monomer \citep[e.g.][]{Sorensen2001}. Thus, one can speculate that, while the transmission spectra may be similar in the two scenarios, the emission spectra should be different and observationally distinguishable.  

\subsubsection{Patchy aerosols}\label{sec:patchy_discuss}

Non uniform aerosol distributions have been studied before in the context of ``patchy'' aerosols. In this formalism different lines of sight see different levels of aerosols; however, variations along individual lines of sight are not taken into account \citep[e.g.][]{Line2016}. In this formalism, transmission spectra become a linear combination of a clear atmosphere and one with aerosols, allowing the transit depth (assuming homogeneously mixed absorbers, $l=0$, and $H\ll R_p$) to be written as:
\begin{equation}
    \frac{\Delta R_T}{H} = f\log\left(\frac{\sigma_{g,\lambda}+\sigma_{a,\lambda}}{\sigma_{g,\lambda_{\rm ref}}+\sigma_{a,\lambda_{\rm ref}}}\right)+(1-f)\log\left(\frac{\sigma_{g,\lambda}}{\sigma_{g,\lambda_{\rm ref}}}\right)
\end{equation}
where $f$ is the fraction of the lines of sight with aerosols and $\sigma_{g,\lambda}$ is the cross-section of the gas, which in the case where aerosol opacity dominates over the gas opacity along the aerosol patches, we find:
\begin{equation}
    \frac{\Delta R_T}{H} \approx f\log\left(\frac{\sigma_{a,\lambda}}{\sigma_{a,\lambda_{\rm ref}}}\right)+(1-f)\log\left(\frac{\sigma_{g,\lambda}}{\sigma_{g,\lambda_{\rm ref}}}\right) \label{eqn:patchy1}
\end{equation}
Thus, one can see how patchy coverage can mute transmission spectra, resulting in a degeneracy between atmospheric metallicity and the degree of aerosol coverage \citep[e.g.][]{Line2016}. Comparison of Equation~\ref{eqn:patchy1} with Equation~\ref{eqn:transit_depth1} implies that the total effective extinction cross-section in the patchy aerosol formalism is given by:
\begin{equation}
    \sigma_{\rm eff,patchy}\approx \sigma_{a,\lambda}^f\sigma_{g,\lambda}^{1-f}
\end{equation}
which appears mathematically similar to the spectra derived from stacking a fraction $f$ aerosol laden exoplanets with $1-f$ aerosol free exoplanets \citep[e.g.][]{Kirk2025}. We can compare the patchy case to the total effective extinction in the clumpy case of:
\begin{equation}
    \sigma_{\rm eff, clumpy}= \frac{3f_{\rm int}}{4N_c} + \sigma_{\rm g,\lambda}
\end{equation}
Thus, a patchy formalism is clearly distinct from our clumpy formalism. In particular, at best, the patchy aerosol formalism will either reproduce the Rayleigh slope in the NIR to MIR for small particles in the limit $f\rightarrow1$, or require large particles with implausibly high aerosol production rates (which are even larger than the homogeneous case, given their partial coverage). The reason for this is that the patchy formalism explicitly neglects variations along the line-of-sight and, therefore, cannot capture the intra-clump absorption, which reduces the effective cross-section's wavelength dependence and the scattering of clump surfaces along with the aerosol free gaps, which increases the mean intensity in the medium. We demonstrate the difference explicitly by comparing the patchy approach with the clumpy approach in Figure~\ref{fig:patchy_compare}. 

\begin{figure}
    \centering
    \includegraphics[width=\columnwidth]{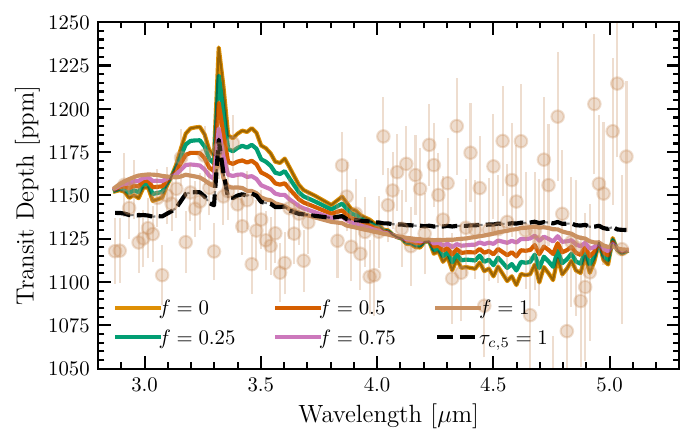}
    \caption{Modelled transmission spectra for TOI-776c computed in the patchy framework for different covering fractions ($f$), with aerosol free ($f=0$) and full homogeneous aerosol coverage ($f=1$) shown as end-members. We also show the case for our clumpy aerosol distribution with $\tau_{c,5}=1$. All models use aerosol production rates of $10^{-10}$~g~cm$^{-2}$~s$^{-1}$ with aerosol particles 0.1$\mu$m in radius. As detailed in the text, patchy aerosol coverage can only mute gas-free spectra and tend to the homogeneous aerosol case as $f\rightarrow 1$. Therefore, patchy aerosols cannot reach the same spectral flatness as clumpy aerosols without the use of unphysically large particles.}
    \label{fig:patchy_compare}
\end{figure}

Given the detection of asymmetric morning and evening terminators with JWST \citep[e.g.][]{Espinoza2024}, which has been interpreted in terms of differences in aerosols coverage on the morning and evening terminators \citep[e.g.][]{Fu2025}, a patchy formalism is likely required at some-level. In fact, the clumpy and patchy formalism is not mutually exclusive. The clumpy formalism can be implemented on top of the patchy formalism, similar to the case of Jupiter, where banded clouds show substructure \citep[e.g.][]{Matcheva2005}.

\subsection{Clumpy aerosols in the wider context}

In this work, we have focussed on TOI-776c, given it has an observed flat JWST spectrum and a Lyman-$\alpha$ transit demonstrating a primordial atmosphere with a metallicity below a few hundred. However, TOI-776c is the archetype of a growing population of planets. Indeed, TOI-776b, 776c's inner companion, also presents with a flat JWST spectrum \citep{Alderson2025} and also has a Lyman-$\alpha$ transit (although, due to its closer proximity to the star, its transit is weaker and thus less constraining \citealt{Owen2023}). Furthermore, other JWST targets with flat spectra show helium absorption. Unlike hydrogen, which could be produced in principle from photo-dissociation of a steam atmosphere (which is ruled out for TOI-776c), helium is a tracer of a primordial atmosphere. However, due to uncertainties in modelling helium transits, it is harder to use them to truly confirm a planet's low metallicity nature, although recent modelling suggests it might be sensitive to metallicity \citep[e.g.][]{Zhang2025}. However, within the wider context, it certainly appears that a significant subset, if not the majority, of sub-Neptunes have primordial H/He-dominated atmospheres, and as such, high-altitude aerosols that behave in a grey manner are required to explain their transmission spectra. 

There is likely a variety in the composition of sub-Neptunes. Phase curve measurements of GJ1214b are consistent with a hazy, high-metallicity atmosphere, albeit one that is inferred to have unusually reflective hazes \citep{Kempton2023}. \citet{Gao2023} used microphysical models to demonstrate that the flatness of the transmission spectrum over a broad range of wavelengths (1–12~$\mu$m) required extremely high metallicities (in excess of 1000, by number). Thus, it is possible that the inclusion of clumpy aerosols would relax this metallicity constraint; however, as we found in our calculations, this does not remove the requirement for a high haze production rate. 

Several sub-Neptunes have shown molecular features in their spectra, leading to a variety of different conclusions. \citet{Piaulet2024} detected water features in GJ 9827d with small amplitudes. When combined with the lack of exospheric detection in helium or hydrogen, they argued for a high-metallicity water-world scenario. \citet{Madhusudhan2023} detected large-amplitude methane features in K2-18b, a planet that also shows a tentative Lyman-$\alpha$ transit \citep{dosSantos2020}, concluding that its observable atmosphere was H/He-dominated. Similarly, TOI-270d's spectrum contained a rich number of molecular features, again leading to the conclusion that the observable atmosphere was H/He-dominated \citep[e.g.][]{Benneke2024,Holmberg2024}. Likewise, water features in the JWST transmission spectrum of TOI-421b with large amplitudes also indicated a low-metallicity H/He-dominated observable atmosphere \citep[e.g.][]{Davenport2025}. Despite this observational progress, there is still considerable debate about how to interpret the composition of a sub-Neptune's observable atmosphere in the context of its envelope composition and, therefore, its origins \citep[e.g.][]{Rigby2024,Luu2024,Shorttle2024,Hu2025,Rogers2025,Heng2025}. 

Thus, while there is clearly diversity in the atmospheric composition of sub-Neptunes, observations at temperature ranges where hydrocarbon hazes are less efficiently produced have indicated that low mean-molecular-weight H/He-dominated atmospheres exist \citep{Davenport2025}. Coupling this with the fact that the larger sub-Neptunes must possess voluminous H/He-dominated envelopes to explain their densities \citep[e.g.][]{JontofHutter2016} suggests that many of the flat-spectrum sub-Neptunes should also similarly be H/He-dominated atmospheres. Therefore, clumpy aerosol distributions could be the solution to explain these flat spectra out to MIR wavelengths without appealing to unphysically large particle sizes and high production rates. This is especially true in the context of planets that exhibit both flat spectra and escaping primordial atmosphere detections, such as TOI-776b/c \citep[e.g.][]{Alderson2025,Teske2025,Loyd2025}, TOI-836c \citep[e.g.][]{Wallack2024,Zhang2025} and GJ 3090b  \citep[e.g.][]{Ahrer2025}. 

Therefore, given the importance of studying sub-Neptunes, more work on clumpy aerosol distributions in exoplanet atmospheres is warranted. In particular, identifying physical mechanisms that might lead to clumping, as well as studying emission spectra to observationally distinguish them from alternative models, is critical.

\section{Summary}\label{sec:summary}

Understanding the nature of sub-Neptunes has been hampered by recent JWST results, which reveal that their transmission spectra are often flat. Explaining these flat spectra typically requires either a high metallicity, high mean molecular weight atmosphere or high-altitude grey aerosols. However, the detection of escaping hydrogen and helium from some sub-Neptunes points to primordial hydrogen/helium-dominated atmospheres, indicating that high-altitude aerosols are responsible, rather than high mean molecular weight atmospheres. Standard aerosol models with homogeneous distributions struggle to reproduce flat, grey spectra: small particles absorb longer-wavelength radiation much less efficiently than shorter wavelengths, producing spectral slopes in transmission, whereas large particles, which act as grey absorbers at NIR to MIR wavelengths, settle out from high altitudes too quickly.  

In this work, we explored heterogeneous ``clumpy'' aerosol distributions. Our main results are as follows:

\begin{enumerate}
    \item Clumpy aerosol distributions have effective extinction opacities that are, on average, lower (due to the creation of optically thin pathways) but exhibit weak or no wavelength dependence, even with small particles. The lack of wavelength dependence arises because photons are either trapped inside individual optically thick clumps, irrespective of their wavelength, or do not interact with a clump. 

    \item Once individual clumps become marginally optically thick ($\tau \gtrsim 0.3$) at 5~$\mu$m, flat JWST 1–5~$\mu$m transmission spectra can be produced.

    \item If the clumps become too optically thick, the covering fraction of the aerosols becomes so low that the transmission spectrum begins to resemble an aerosol-free spectrum. Thus, to flatten transmission spectra, marginally to moderately optically thick clumps are required.

    \item By modelling the JWST G395H transmission spectrum of TOI-776c (a flat-spectrum sub-Neptune with a large hydrogen exosphere), we find that small ($\lesssim 0.1~\mu$m) tholin-like aerosol particles can explain the observed spectrum with individual clump optical depths between $\sim 0.3$ and 10, even with a solar metallicity atmospheric composition.

    \item Unlike homogeneous aerosol distributions, small particles are more effective at flattening transmission spectra in a clumpy distribution because they settle more slowly, making it easier to form high-altitude, optically thick aerosol layers.

     \item In the framework of clumpy aerosol distributions, NIR JWST transmission spectra are completely degenerate between the clump optical depth and the atmospheric metallicity, with moderately optically thick clumps and a low metallicity atmosphere consistent with flat spectra, as well as aerosol-free high metallicity atmospheres; although exospheric detections can be used to put an upper bound on the metallicity.

    \item Since the average extinction decreases and the mean radiative intensity increases with a clumpy aerosol distribution, emission spectra should be modelled and explored as a potential observational test of the clumpy aerosol hypothesis. 
\end{enumerate}

While clumpy aerosol distributions can naturally produce flat transmission spectra, our modelling has been purely phenomenological. Further theoretical work is required to identify the physical mechanisms capable of generating clumpy aerosols distributions in sub-Neptune atmospheres.

\section*{Acknowledgements}
We are grateful to the anonymous referee for comments which improved the manuscript. JEO is supported by a Royal Society University Research Fellowship. JK acknowledges financial support from Imperial College London through an Imperial College Research Fellowship grant. This project has received funding from the European Research Council (ERC) under the European Union’s Horizon 2020 research and innovation programme (Grant agreement No. 853022). This work benefited from the 2024 Exoplanet Summer Program in the Other Worlds Laboratory (OWL) at the University of California, Santa Cruz, a program funded by the Heising-Simons Foundation. We are grateful to Sarah Moran for discussions surrounding aerosol distributions in sub-Neputnes.

\section*{Data Availability}

 The data underlying this article will be shared with reasonable request to the corresponding author.



\bibliographystyle{mnras}
\bibliography{main} 




\appendix

\section{Soot aerosols}\label{sec:soot}
In Section~\ref{sec:discuss} we saw that the opacity window in tholin-like aerosols around 1-2.5~$\mu$m led to a reduced transit depth in the clumpy cloud scenario. Here we demonstrate that this effect is unique to the tholin opacity window in Figure~\ref{fig:soot}, where we remake Figure~\ref{fig:long_wav} with soot aerosols, using the optical constants from \citet{Lavvas2017}. 

\begin{figure*}
    \centering
    \includegraphics[width=\textwidth]{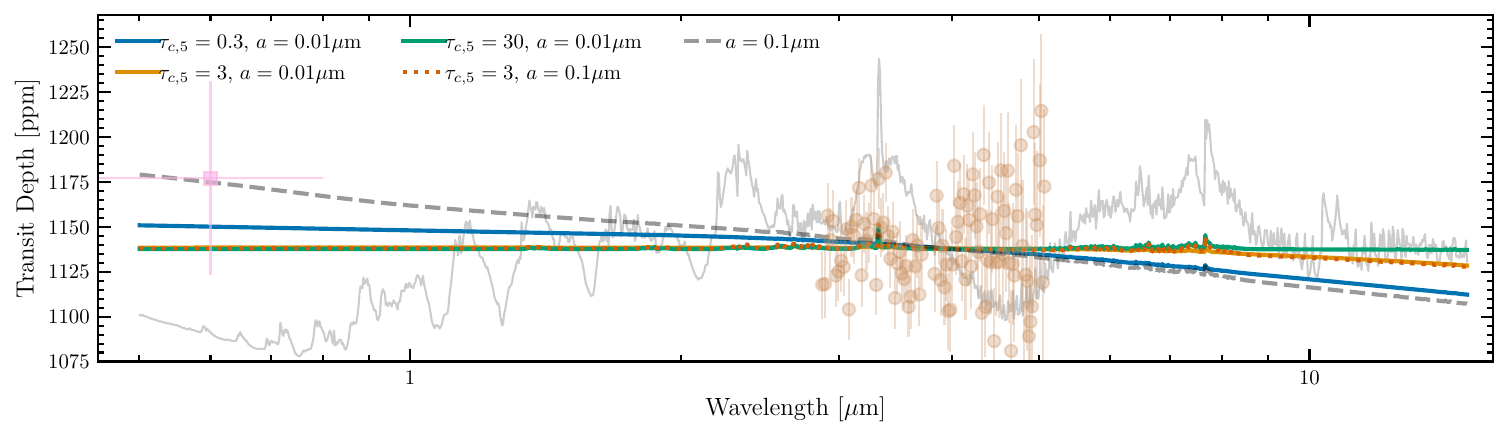}
    \caption{Same as Figure~\ref{fig:long_wav} in the main text that used tholin-like hazes, but instead using soot-like haze particles.}
    \label{fig:soot}
\end{figure*}

\bsp	
\label{lastpage}
\end{document}